\documentclass[useAMS,usenatbib]{mn2e}
\usepackage{psfrag,graphicx}
\usepackage{color}
\usepackage{txfonts}
\usepackage{breqn}
\usepackage{graphicx}
\title[Black hole formation in the early universe]
   {The characteristic black hole mass resulting from direct collapse in the early universe}

\author[Latif et al.]
  {M.~A.~Latif,$^1$
  D.~R.~G.~Schleicher,$^1$ 
  W.~Schmidt,$^1$
  J.~C.~Niemeyer$^1$
   \newauthor 
   $^1$ Institut f\"ur Astrophysik, Georg-August-Universit\"at, \\
    Friedrich-Hund-Platz 1, D-37077 G\"ottingen, Germany}
\date{today}

\pagerange{\pageref{firstpage}--\pageref{lastpage}} \pubyear{2009}

\def\LaTeX{L\kern-.36em\raise.3ex\hbox{a}\kern-.15em
      T\kern-.1667em\lower.7ex\hbox{E}\kern-.125emX}

\begin{document}

\bibliographystyle{mn2e}

\label{firstpage}

 \maketitle
%


\begin{abstract}
{Black holes of a billion solar masses are observed in the infant universe a few hundred million years after the Big Bang. The direct collapse of protogalactic gas clouds in primordial halos with $\rm T_{vir} \geq 10^{4}~K$ provides the most promising way to assemble massive black holes. In this study, we  aim to determine the characteristic mass scale of seed black holes and the time evolution of the accretion rates resulting from the direct collapse model. We explore the formation of supermassive black holes via cosmological large eddy simulations (LES) by employing sink particles and following their evolution for twenty thousand years after the formation of the first sink. As the resulting protostars were shown to  have cool atmospheres in the presence of strong accretion, we assume here that UV feedback is negligible during this calculation. We confirm this result in a comparison run without sinks. Our findings show that black hole seeds with characteristic mass of $\rm 10^{5}~M_{\odot}$ are formed in the presence of strong Lyman Werner flux which leads to an isothermal collapse. The characteristic mass is a about two times higher in LES compared to the implicit large eddy simulations (ILES). The accretion rates increase with time and reach a maximum value of 10 $\rm M_{\odot}/yr$ after $\rm 10^{4}$ years. Our results show that the direct collapse model is clearly feasible as it provides the expected mass of the seed black holes.
}
 
\end{abstract}




\section{Introduction}

Black holes of a few millions to billions solar masses dwell in the center of present day galaxies \citep{Kormendy95, Tremaine02,2011Natur.480..215M,2012Natur.491..729V}. These supermassive black holes (SMBHs) are not only present in the local Universe but have been observed at $\rm z> 6$ \citep{2003AJ....125.1649F,2006AJ....131.1203F,2011Natur.474..616M}. Their formation in the first billions years after the Big Bang is still an open question. 

Numerous models have been proposed to explain the origin and formation of supermassive black holes \citep{1984ARA&A..22..471R,2009MNRAS.tmp..640R,2008arXiv0803.2862D,2009ApJ...702L...5B,2009MNRAS.396..343R,2009ApJ...696.1798T,2010A&ARv..18..279V,2012arXiv1203.6075H,2012ApJ...750...66J,2013ApJ...771..116J}. They include the collapse of dense stellar cluster due to relativistic instability \citep{2004Natur.428..724P,2008ApJ...686..801O, 2009ApJ...694..302D}, remnants of Population III stars \citep{2004ApJ...613...36H} and the direct collapse of protogalactic gas clouds \citep{2002ApJ...569..558O,2003ApJ...596...34B,2006ApJ...652..902S,2006MNRAS.370..289B,2006MNRAS.371.1813L,2008MNRAS.391.1961D,2008arXiv0803.2862D,2010MNRAS.402.1249S,2010MNRAS.tmp.1427J,2010ApJ...712L..69S,2011MNRAS.411.1659L,2013arXiv1304.1369C,2013ApJ...774...64W}. Although, the formation of SMBHs from stellar mass black holes may appear as the most natural way, the feedback from the stars creates a hindrance as they have to accrete at an Eddington rate all the time to reach the observed masses \citep{2007MNRAS.374.1557J, 2008arXiv0811.0820A,2012ApJ...756L..19W}. The masses of black holes resulting from the collapse of dense stellar clusters are relatively low with $\rm \sim 10^{3}~M_{\odot}$ \citep{2009ApJ...694..302D}. On the other hand, the direct collapse seems the most plausible way to assemble SMBHs to provide higher mass black hole seeds.

The formation of SMBHs via direct collapse requires the suppression of fragmentation and efficient accretion of the gas onto the central object. Massive primordial halos of $\rm 10^{7}-10^{8}~M_{\odot}$ irradiated by strong Lyman Werner background fluxes are the most plausible candidates \citep{2001ApJ...546..635O,2007MNRAS.374.1557J,2008MNRAS.391.1961D,2010MNRAS.402.1249S,2011MNRAS.410..919J,2010ApJ...712L..69S,2011MNRAS.418..838W,2011A&A...532A..66L,2012MNRAS.425.2854A,2013MNRAS.tmp..551L}, see also \cite{2012MNRAS.422.2539I,2013A&A...553L...9V} for alternatives. Numerical simulations support this scenario and show that in the presence of strong Lyman Werner flux an isothermal collapse yields massive objects   \citep{2003ApJ...596...34B,2008ApJ...682..745W,2009MNRAS.393..858R,2011MNRAS.411.1659L,2013MNRAS.433.1607L}. The final fate of these objects is not yet very well understood and depends on the gas mass accretion rates.

Theoretical models \citep{2008MNRAS.387.1649B,2010MNRAS.402..673B,2012ApJ...756...93H,2013arXiv1305.5923S,2013ApJ...768..195W,2013arXiv1308.4457H} propose the formation of supermassive stars (normal stars of higher masses) or quasistars (stars with black holes at the center) as an intermediate stage to the formation of SMBHs. The work by \cite{2013arXiv1305.5923S} suggests that for accretion rates $\rm >0.14~M_{\odot}/yr$, the core of the star collapses into a black hole, forming a so-called quasi-star while lower accretion rates lead to the formation of a supermassive star. \cite{2012ApJ...756...93H,2013arXiv1308.4457H} show that for accretion rates $\rm > 10^{-2}~M_{\odot}/yr$, supergiant stars of $\rm 10^{5}~M_{\odot}$ form via rapid accretion while maintaining the cool atmospheres on the Hayashi track. \cite{2013arXiv1305.5923S} have suggested that this may be true up to a mass scale of $\rm 3.6 \times 10^{8} ~ \dot{m}^{3}~M_{\odot}$, where $\rm \dot{m}$ is the mass accretion rate in units of $\rm M_{\odot}/yr$. These models for the evolution of supermassive stars show that in the presence of accretion rates $\rm > 10^{-2}~M_{\odot}/yr$, such stars have a bloated envelope and lower surface temperatures, inhibiting the emission of ionizing flux. A recent study by \cite{2013MNRAS.433.1607L} reported high accretion rates of $\rm 1~M_{\odot}/yr$ in these halos and the possibility for the formation of massive objects in a short time.

In this article, we for the first time explore the characteristic mass scale of seed black holes and the time evolution of accretion rates in massive primordial halos illuminated by a strong Lyman Werner background UV flux. To accomplish this goal, we perform high resolution cosmological large eddy simulations to ensue the collapse below parsec scales by exploiting the adaptive mesh refinement method and employing sink particles to follow the accretion for longer time scales. To verify our results, we also perform a comparison run without sink particles. We evolve the simulations for twenty thousand years after the formation of the first sink and employ a Jeans resolution of 32 cells to resolve turbulent eddies. The subgrid-scale turbulence model of \cite{SchmNie06b} is used to take into account unresolved turbulence. This study will enable us to compute the masses of seed black holes formed in the massive primordial halos and test the feasibility of a direct collapse model. 

This article is organized in the following way. In section 2, we describe the numerical methods and simulations setup. We present our results in section 3. In section 4, we summarize the main findings of this study and discuss our conclusions.

\section{Computational methods}
\subsection {Simulation Setup}
The simulations presented here are performed using a modified version of the Enzo code \citep{2004astro.ph..3044O,2013arXiv1307.2265T}. Enzo is a Eulerian grid based, massively parallel, cosmological adaptive mesh refinement code. The hydrodynamical equations are solved using a 3rd order accurate piece-wise parabolic method (PPM). The dark matter dynamics is solved using the particle mesh technique.

The initial simulation setup is the same as in our previous study \citep{2013MNRAS.433.1607L}. Here, we present a short summary of the initial conditions and the simulations setup. Our simulations are started with cosmological initial conditions at $\rm z=100$. We use the publicly available inits package to generate nested grid initial conditions. These simulations were first run with a uniform grid resolution of $\rm 128^{3}$ and the dark matter haloes are selected at $\rm z=15$. The simulated volume has a comoving size of $\rm 1~Mpc~h^{-1}$ and the most massive halo lies at the center of the box. We further employ two initial nested refinements levels with a resolution of $\rm 128^{3}$ cells each and initialize 5767168 particles to simulate the dark matter dynamics. In the central 62 kpc region of the halo, we allow 15 dynamical refinement levels during the course of the simulations which yields an effective resolution of sub parsec scales (10,000 AU in comoving units). We resolve the Jeans length by 32 cells during the entire course of the simulations. Once the maximum refinement level is reached, we employ sink particles to follow the evolution for twenty thousand years after the formation of the first sink. 

We employ a strong Lyman Werner flux of strength $\rm 10^{3}$ in units of $\rm J_{21}$ produced by the first generation of stars with radiation spectra of $\rm 10^{5}~K$ \citep{2001ApJ...546..635O,2007MNRAS.374.1557J,2008MNRAS.391.1961D,2010ApJ...712L..69S,2011MNRAS.418..838W,2011A&A...532A..66L,2012MNRAS.426.1159S,2012MNRAS.425.2854A}. To follow the thermal evolution, we self consistently solve the rate equations of $\rm H$, $\rm H^{+}$, $\rm He$, $\rm He^{+}$,~$\rm He^{++}$, $\rm e^{-}$,~$\rm H^{-}$,~$\rm H_{2}$,~$\rm H_{2}^{+}$ in the cosmological simulations. We ignore the effect of self-shielding as our main focus is on $\rm H_{2}$ free halos.  

To take into account the unresolved turbulence, we use the subgrid scale turbulence model proposed by \cite{SchmNie06b}. The adaptively refined large eddy simulations technique is used to apply the SGS model in cosmological AMR simulations \citep{2009ApJ...707...40M}. We perform large eddy simulations (LES) and compare our results with implicit large eddy simulations (ILES). The approach of LES is based on scale separation mechanism i.e., separation of the resolved and unresolved scales, and connect them through an eddy-viscosity closure for the transfer of energy between the grid scales. The turbulent viscosity is computed from the grid scale and the SGS turbulence energy, i.~e., the kinetic energy associated with numerically unresolved turbulent velocity fluctuations. On the other hand, ILES use only the numerical dissipation produced from the discretization errors of fluid dynamics equations. 

In all, we have performed 6 simulations (3 LES, 3 ILES) with sink particles for three distinct halos A, B and C and a comparison run without sinks particles for halo A. The halo properties are given in the table 1 of \cite{2013MNRAS.433.1607L}. Their collapse redshifts are 12.6, 10.8 and 13.6 respectively. For a comparison run without sinks, simulations are evolved adiabatically after they reach the maximum refinement level. This method allows us to follow the collapse for a longer times.

\subsection {Sink Particles}
The need to resolve the Jeans length to stellar densities and Courant constraints on the calculation of the time step make it impossible to follow the collapse to the smallest scales while evolving the simulations for a long time. Therefore, sink particles are introduced to represent the gravitationally bound objects undergoing a free-fall collapse. This approach has been successfully employed in both SPH as well as in AMR codes \citep{1995MNRAS.277..362B,2004ApJ...611..399K,2010ApJ...713..269F}. Here, we employ the sink particle algorithm by \cite{2010ApJ...709...27W} to represent the protostars. Sinks are created when a grid cell is at the maximum refinement level, the density exceeds the Jeans density (i.e., it violates the Truelove criterion) and the overall flow is convergent, which is implicitly covered by a density threshold check. Furthermore, particles are merged if they are created within the accretion radius. The initial mass of the sinks is calculated such that the cell is Jeans stable after the subtraction of the sink mass. The initial velocity of the sinks is computed based on momentum conservation. Further details of the sinks algorithm can be found in the reference article \citep{2010ApJ...709...27W}.  

The sink particles in our case do not accrete gas directly from the grid as typically computed using the Bondi-Hoyle accretion. This effect is compensated by allowing the formation of additional sinks and subsequently, they are merged if they are formed within the accretion radius which we choose as the Jeans length. Using sink particles, we follow the collapse for longer dynamical time scales. This approach enables us to compute the masses of seed black holes and the time evolution of mass accretions.

\begin{figure*}
\vspace{-1.0cm}
\hspace{-9.0cm}
\centering
\begin{tabular}{c}
\begin{minipage}{8cm}
\includegraphics[scale=0.8]{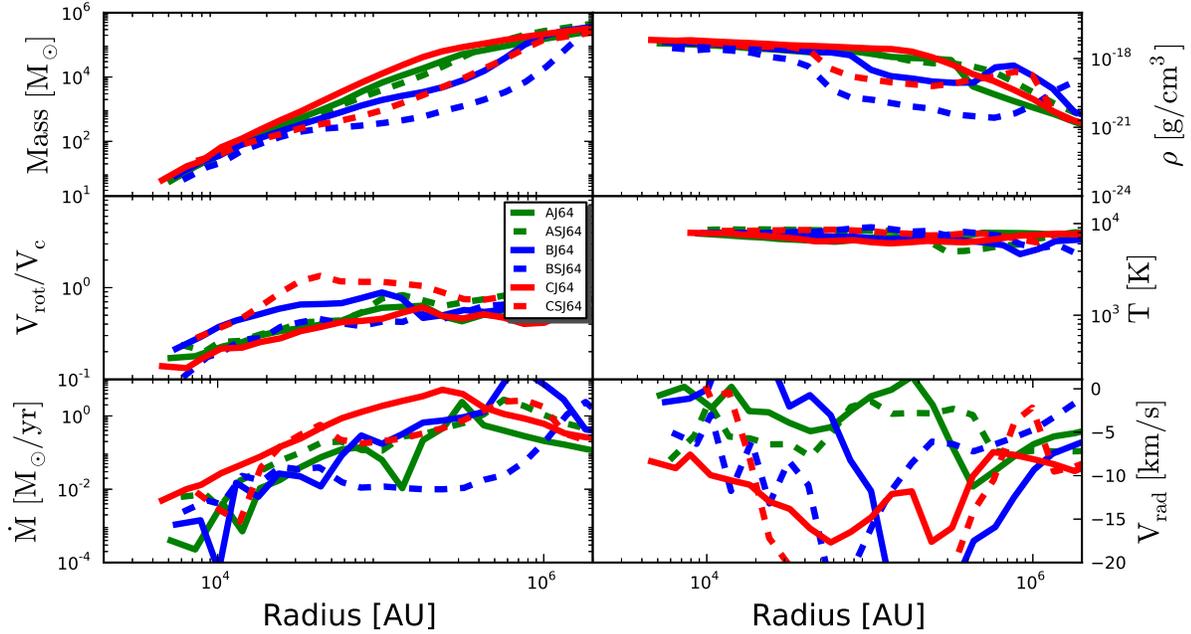}
\end{minipage}
\end{tabular}
\caption{ Radially binned spherically averaged radial profiles are shown for the halos A, B and C. The solid lines represent ILES runs while the dashed lines stand for LES runs. Top left and right panels show the enclosed mass and density radial profiles. The rotational velocity and temperature radial profiles  are depicted in the middle left and right panels. The bottom panels show accretion rates and radial velocity profiles.}
\label{figh4}
\end{figure*}

\begin{figure*}
\centering
\begin{tabular}{c}
\begin{minipage}{6cm}
 \hspace{-5cm}
\includegraphics[scale=0.3]{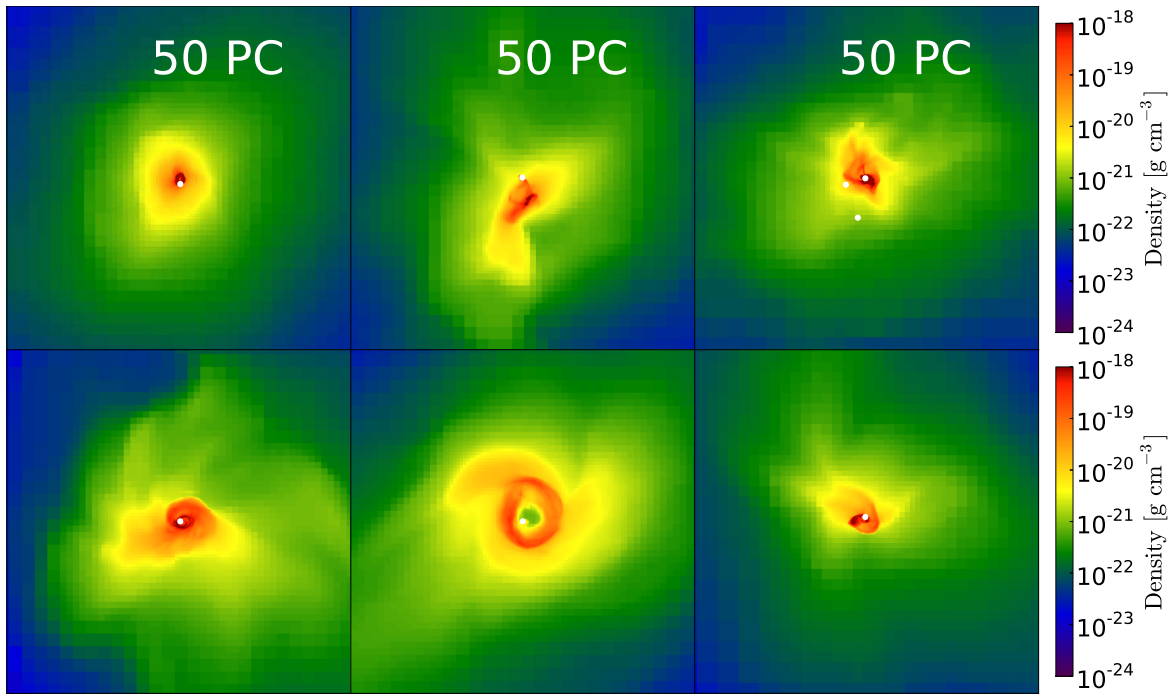}
\end{minipage}
\end{tabular}
\caption{Density projections are shown for the central 50 pc twenty thousand years after the formation of the first sink. The white spheres represent sink particles and are overplotted on the density projections. They have typical masses of $\rm \sim 10^{5}~M_{\odot}$.}
\label{fig1}
\end{figure*}

\begin{figure*}
\centering
\begin{tabular}{c}
\begin{minipage}{6cm}
 \hspace{-5cm}
\includegraphics[scale=0.3]{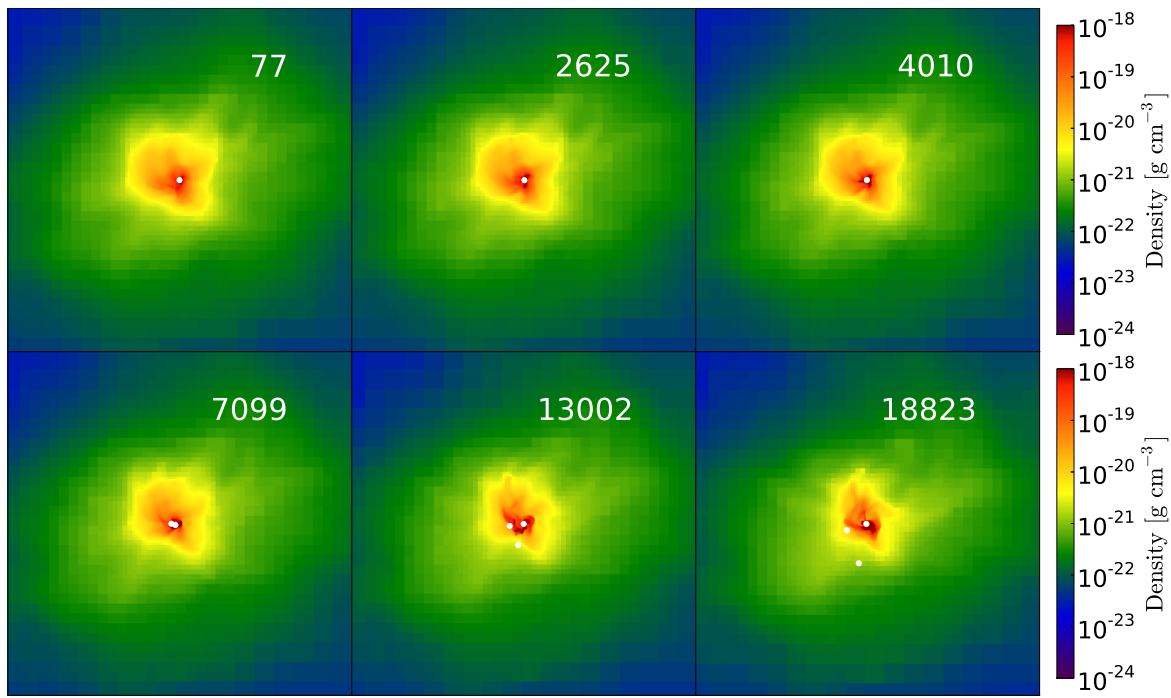}
\end{minipage}
\end{tabular}
\caption{ Time evolution of the density projections is shown in this figure for an ILES run (halo C). The time in years, after the formation of the first sink is shown in each projection. The white spheres represent sink particles and are overplotted on the density projections. They have typical masses of $\rm \geq 10^{4}~M_{\odot}$.}
\label{fig2}
\end{figure*}

\begin{figure*}
\centering
\begin{tabular}{c c}
\begin{minipage}{4cm}
\hspace{-4cm}
\includegraphics[scale=0.4]{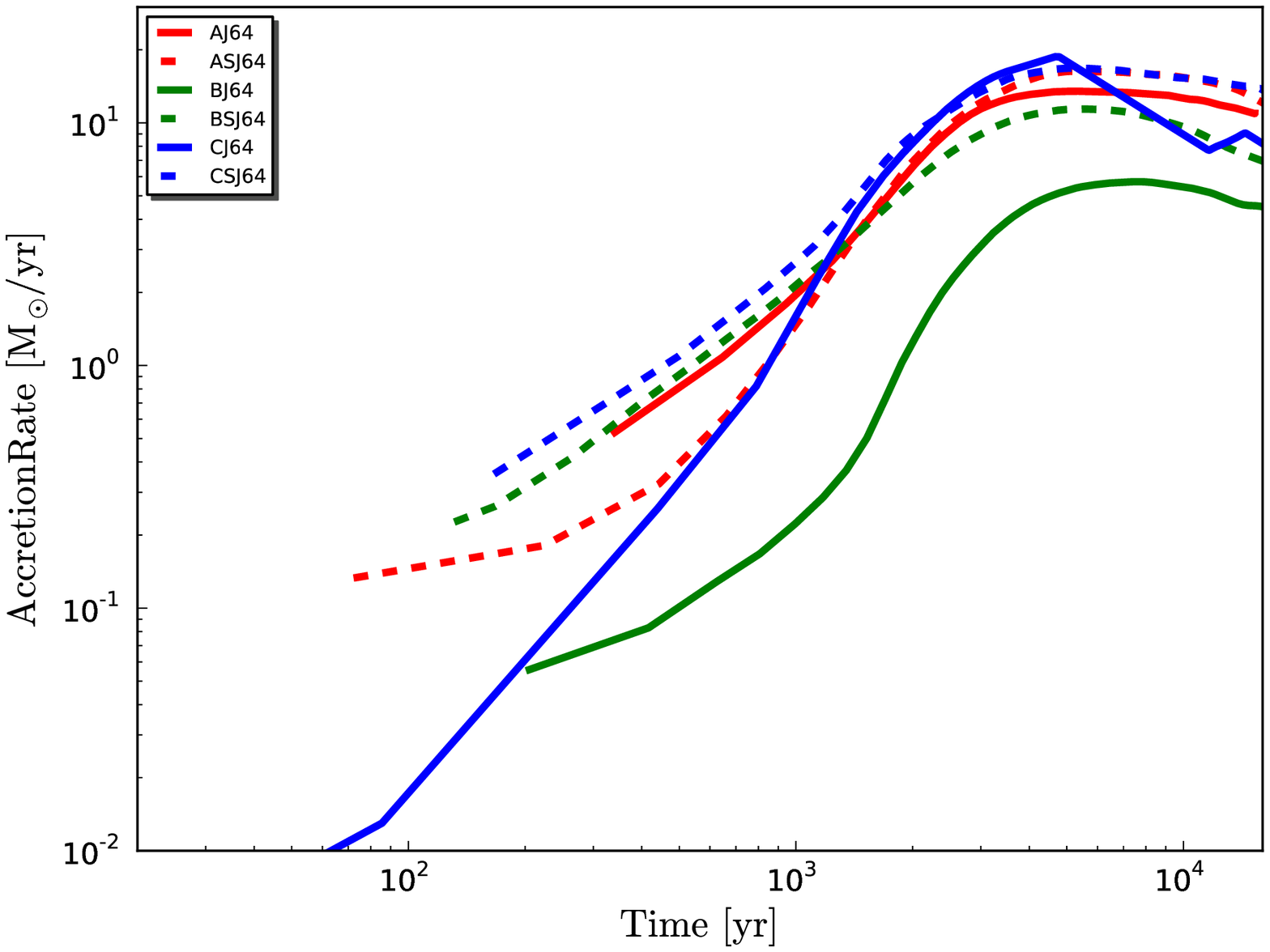}
\end{minipage} &
\begin{minipage}{4cm}
\includegraphics[scale=0.4]{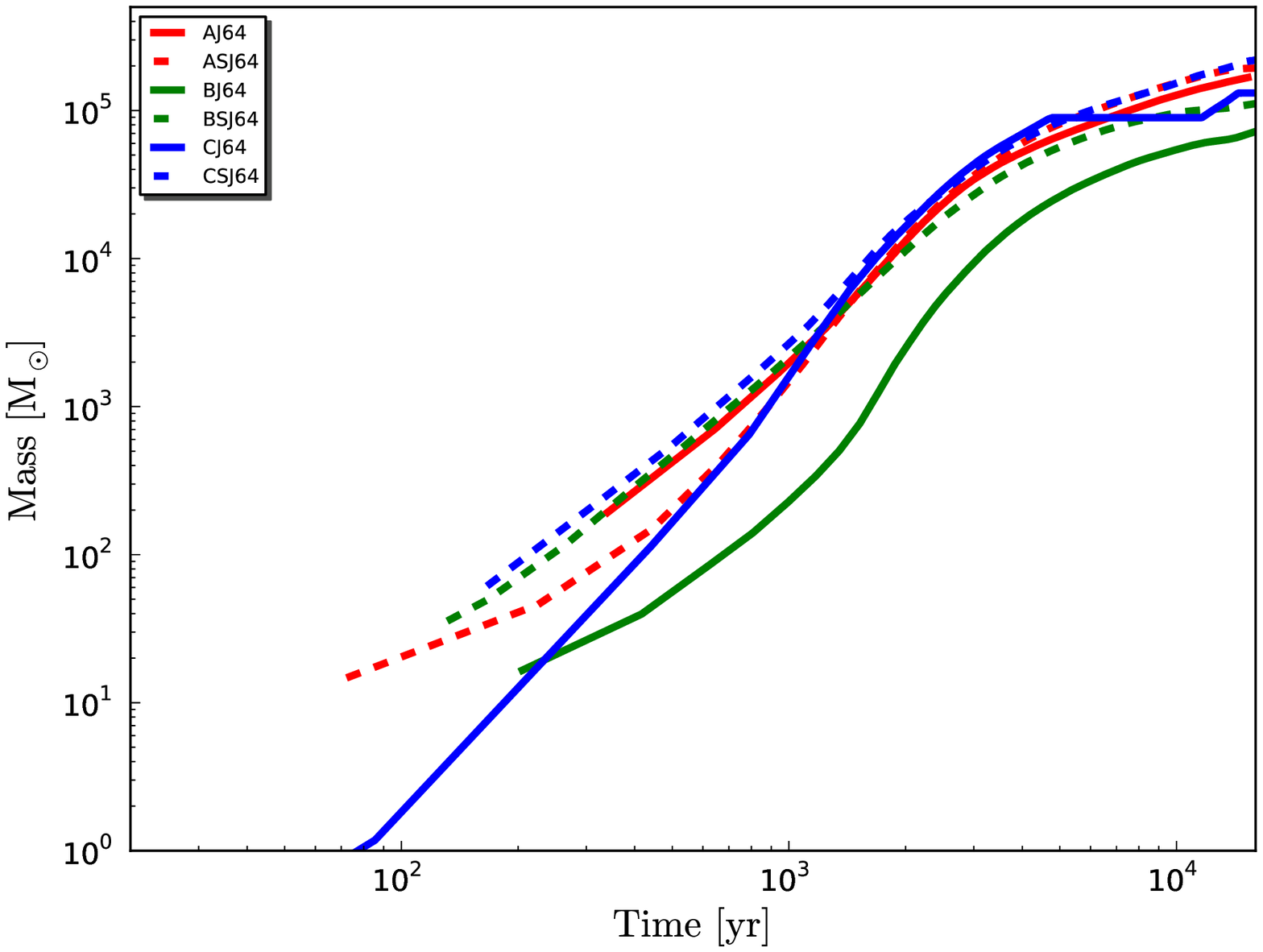}
\end{minipage} 
\end{tabular}
\caption{Time evolution of the accretion rates on the most massive sinks is shown in the left panel of this figure. The time evolution of the masses of the most massive sinks is depicted in the right panel. The solid lines represent ILES runs while dashed lines stand for LES runs.}
\label{fig5}
\end{figure*}

\section{Main Results}
\subsection{Simulations with sink particles}
We have performed 6 cosmological simulations with sink particles (3 LES and 3 ILES) for three distinct haloes  A, B and C and employing a constant Lyman Werner background UV flux of $\rm J_{21}=10^{3}$ for stellar spectra of $\rm 10^{5}$ K.  We followed the evolution for twenty thousand years after the simulations reached the maximum refinement level and determined the characteristic mass scale of the most massive objects.  The results obtained from the cosmological large eddy simulations are presented here. After the simulations are started at $\rm z=100$, massive halos are formed around redshift 18, and gas falls into the dark matter potentials and gets shock heated.

The general properties of the halos, twenty thousand years after the formation of the first sink are shown in figure \ref{figh4}. The formation of molecular hydrogen remains suppressed in the presence of strong $\rm H_{2}$ photodissociating background and consequently, an isothermal collapse occurs. The central temperature of these halos is about 7000 K. The maximum density in the halos is a few times $\rm 10^{-18}~g/cm^{3}$. There is an initial rise in the density at larger radii and then it becomes almost constant. This trend is observed for all halos. The deviation in the density radial profiles from an isothermal behavior may result both from the removal of the gas by sink particles from the grid as well as the formation of a disk due to the non-zero angular momentum. The typical radial velocities are a few $\rm km/s$ which indicate the relatively high gas infall rates at this stage of the collapse. The radial profile of the mass accretion rates shows that the accretion rate in the surroundings of the halo is about $\rm 1~M_{\odot}/yr$ and decreases down to $\rm 10^{-3}~M_{\odot}/yr$ in the core of halo, the region which corresponds to the Jeans length. This behavior seems to be consistent for all halos. The ratio of rotational to circular velocities is about 1 in the surroundings of the halo and declines towards the center of the scale which corresponds to the Jeans length. It further shows that there is high degree of rotational support in the halo. The mass profiles increase with $\rm r^{2}$, the deviations from this behavior come from the differences in the underlying density structure for various runs. 

The state of the simulations at their collapse redshift is shown in figure \ref{fig1} and is presented by the density projections. The fragmentation in these halos remains suppressed and a single massive sink is formed in all haloes except one ILES run as discussed below. We note that the underlying density distribution is different from halo to halo for both LES and ILES. We attribute these differences to the properties of the halos and  the occurrence of various processes such as the removal of gas by the sinks and turbulent stresses. We further show the time evolution of the density structure for the ILES run (halo C) in figure \ref{fig2} where the formation of multiple sinks is observed. The formation of a second sink takes place about 7000 years after the formation of the first sink. The third sink particle in this case is formed about 10,000 years after the formation of the first sink. The mass of the most massive sink is about $\rm 10^{5}~M_{\odot}$. The masses of two particles are almost comparable and may lead to a binary or even multiple system. Fragmentation in this halo is due to the local compression of the gas by the turbulence. Similarly, fragmentation was observed in our previous study \citep{2013MNRAS.433.1607L} with a different approach employed for the evolution of simulations. Here, no fragmentation is found in LES runs for all simulated halos.

\begin{figure}
\centering
\begin{minipage}{4cm}
\hspace{-2cm}
\includegraphics[scale=0.4]{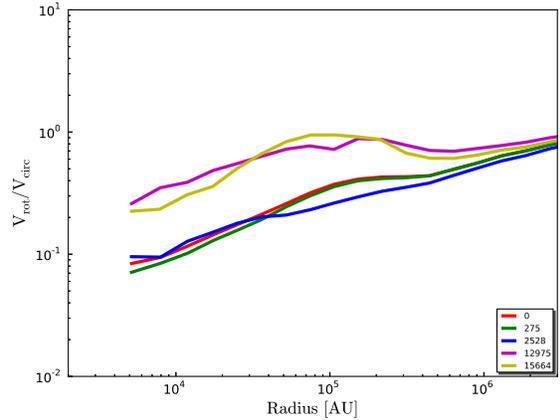}
\end{minipage} 
\caption{The time evolution of the ratio of rotational to Kepler velocity as a function of radius is shown in the figure. The time corresponding to each profiles is shown in the legend and its units are in years. }
\label{figvrot}
\end{figure}

The time evolution of the mass accretion rates for the most massive sinks is shown in figure \ref{fig5} for both LES and ILES. The accretion rate increases with time and reaches a peak value of about 10 solar masses per year in about a time of a few thousand years. Such an accretion rate is in accordance with Bondi-Hoyle accretion 
\begin{equation}
\dot{M}= {4 \pi  \rho G^{2} M^{2} \over c_{s}^{3}}.
\end{equation}
Considering $\rm M= 10^5~M_{\odot}$, $\rm c_{s}=12~km/s$ and $ \rho = 5 \times 10^{-19} g/cm^{3}$, the typical values in our case, the expected accretion rate from the above equation is 9.7 $\rm M_{\odot}/yr$ comparable to the values in our simulations.
After about 10, 000 years the accretion starts to become constant for LES. This time scale can be understood from the infall of a point mass m at distance R from a point source of mass M. The expression for the infall time can be derived from Kepler's third law of motion and is given as 
\begin{equation}
T_{ff}= {\pi \over 2}{R^{3/2} \over \sqrt{2 G (M+m)}}.
\end{equation}
For a point source of mass $\rm 10^5~M_{\odot}$ and a distance of $\rm 3 \times 10^{4}~AU$, the infall time is about 9, 400 years and is in good agreement with our results. This characteristic behavior is noticed for all halos and for both LES and ILES runs. The decline in accretion rates after 10, 000 years in one of the ILES runs is due to the formation of multiple sinks in  halo C and a decrease in the density in the surroundings of the sink particle. The  constant accretion rates after 10, 000 years are a consequence of enhanced rotational support in the halo with time. The ratio of rotational to Keplerian velocity ($\rm v_{rot}/v_{cir}$) increases with time as shown in figure \ref{figvrot}. It indicates that as $\rm v_{rot}/v_{cir}$ increases, it takes longer time for the gas to reach the center of the halo. It is further noted that LES runs lead to higher accretion rates compared to the ILES runs. We attribute the higher accretion rates (at least about a factor of 2 compared to ILES) in LES to the presence of an additional viscosity term. The occurrence of such high accretion rates has important implications for the evolution of supergiant stars. It is expected that in the presence of such high accretion rates the ionizing feedback from supermassive stars remains suppressed \citep{2012ApJ...756...93H,2013arXiv1305.5923S}. 

As pointed out previously, our main aim here is to compute the characteristic mass scale of seed black holes resulting from the direct collapse model. The time evolution of the sink mass in shown in right panel of figure \ref{fig5}. The figure shows that sinks reach masses of $\rm 10^{5}~M_{\odot}$ in about a time of a few times $\rm 10^{4}$ years. The masses of LES runs are higher (at least about a factor of 2) compared to the ILES runs. This is the consequence of the higher accretion rates in LES. It is found that a few percent of the halo mass goes into the sinks. 

\begin{figure}
\centering
\begin{minipage}{4cm}
\hspace{-2cm}
\includegraphics[scale=0.4]{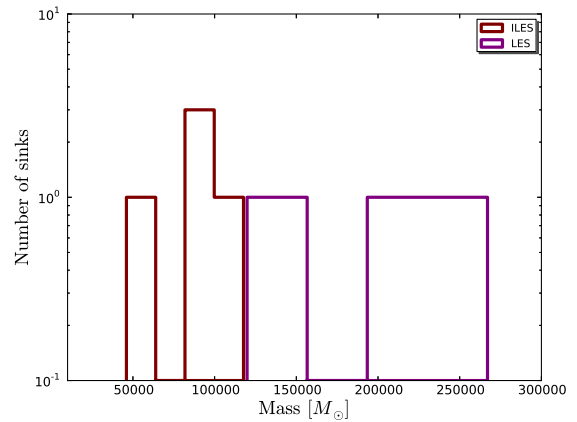}
\end{minipage} 
\caption{Mass distribution of the sinks is shown in this figure. The purple color shows the sink mass distribution for the LES runs while brown color represents ILES runs.}
\label{fig8}
\end{figure}

The mass distribution of the sinks for individual runs is illustrated in figure \ref{fig8}. It is clearly visible from the figure that LES runs produce higher mass sinks compared to the ILES runs.  It can be noted that a single sink particle of $\rm 10^{5}~M_{\odot}$ is formed per halo except for halo C which has two additional particles of about $\rm 5 \times 10^{4}~M_{\odot}$ and $\rm 8 \times 10^{4}~M_{\odot}$. 

\subsection{Comparison run without sinks}
\begin{figure*}
\centering
\hspace{9cm}
\includegraphics[scale=0.8]{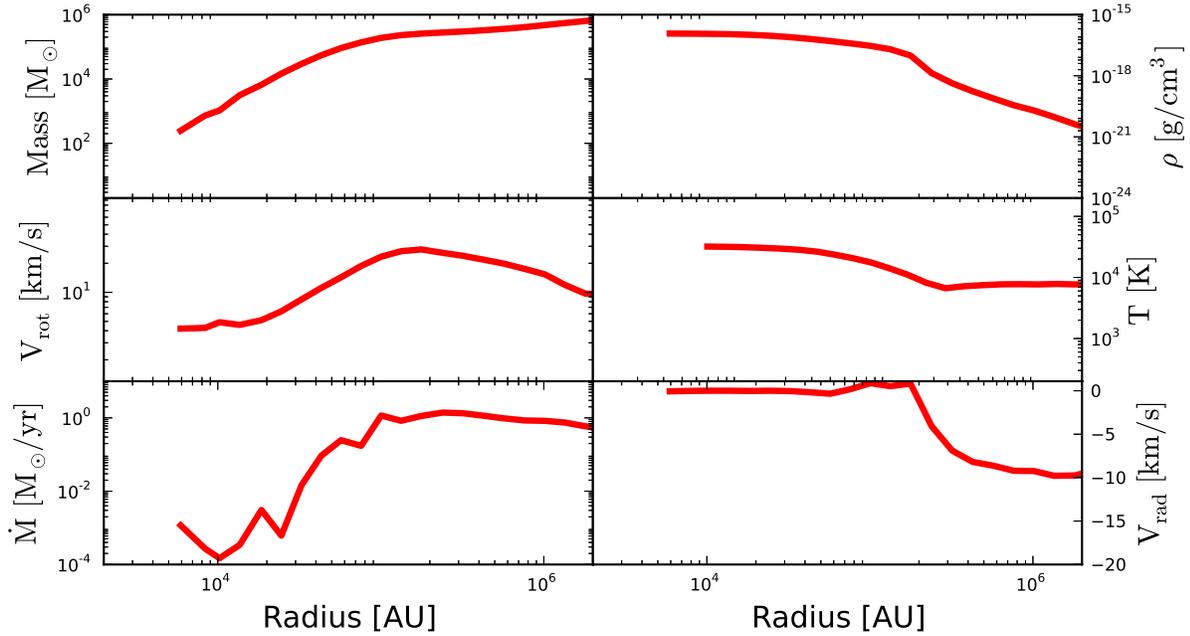}
\caption{Radially binned spherically averaged radial profiles are shown for the run without sinks (for halo A). The radial profiles of  the enclosed mass and the density are shown in the top left and right panels. The middle left panel shows the rotational velocity profile while the right middle panel depicts the temperature radial profiles. The bottom panels show the mass accretion rate and radial velocity profiles.}
\label{fig9}
\end{figure*}

In addition to our 6 cosmological runs with sink particles, we have  performed one cosmological simulation where we evolve the simulation adiabatically at densities above a few times $\rm 10^{-18}~g/cm^{3}$ after reaching the maximum refinement level. The properties of the halo are shown in figure \ref{fig9} after 20,000 years of evolution. The density radial profile shows an isothermal behavior at larger radii and becomes almost flat in the center due to the adiabatic evolution. The temperature is about 8000 K and starts to increase with density to make the collapse stable at the smallest scales. The radial infall velocity is about 10 $\rm km/s$ and becomes almost constant in the center. The mass accretion rate is about 1 $\rm M_{\odot}/yr$  as collapse proceeds on the larger scales and decreases down to the $\rm 10^{-4}~M_{\odot}/yr$ within the Jeans length. The rotational velocity is low in the center, peaks around $\rm 10^{5}~AU$ and declines down as it follows the Keplerian velocity. The mass radial profiles increase with $r^2$ in the center and  becomes flat at larger radii. This indicates that most of the mass lies in the central clump.

\begin{figure*}
\centering
\begin{tabular}{c c}
\begin{minipage}{4cm}
\hspace{-5cm}
\includegraphics[scale=0.32]{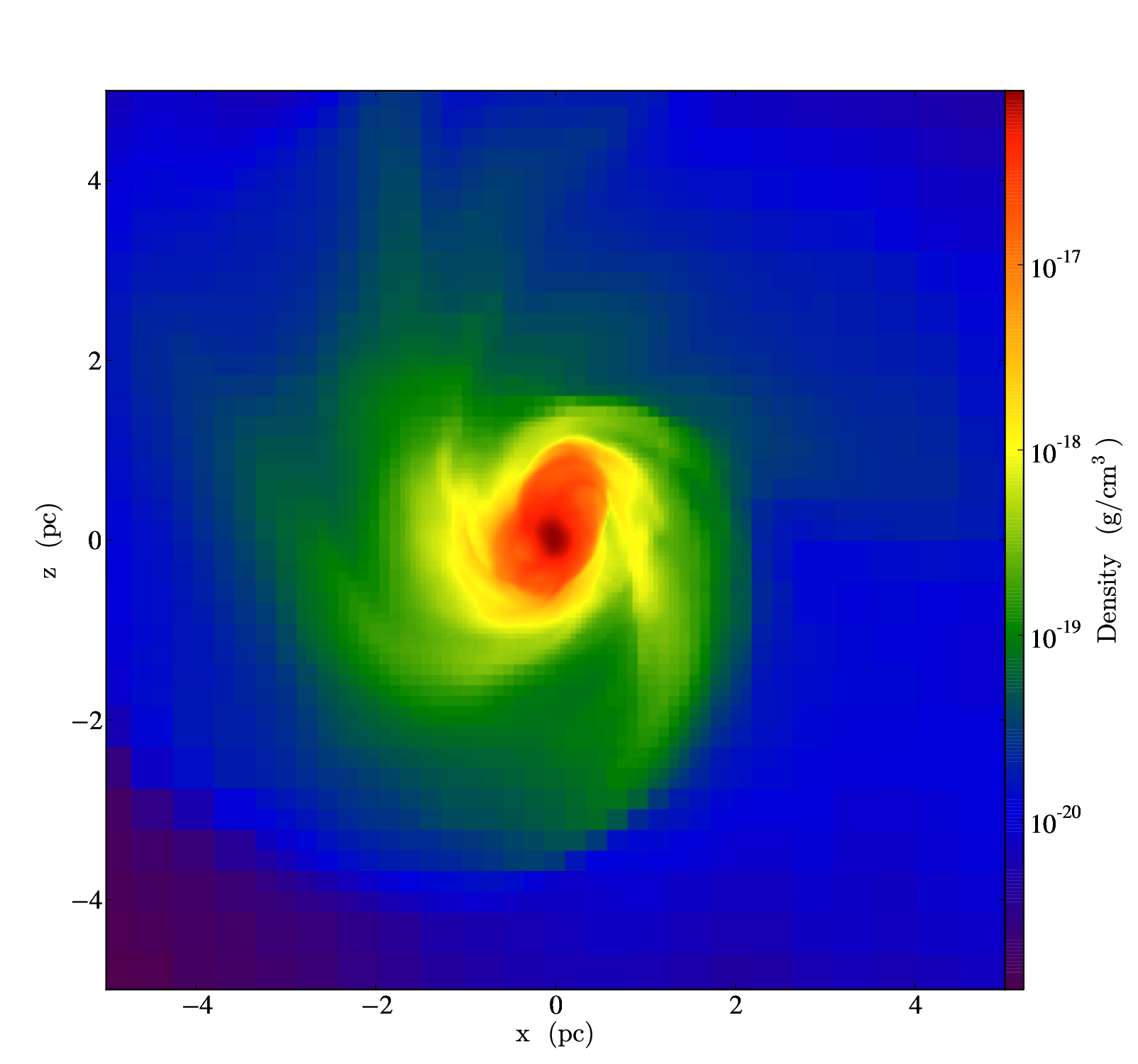}
\end{minipage} &
\begin{minipage}{4cm}
\includegraphics[scale=0.32]{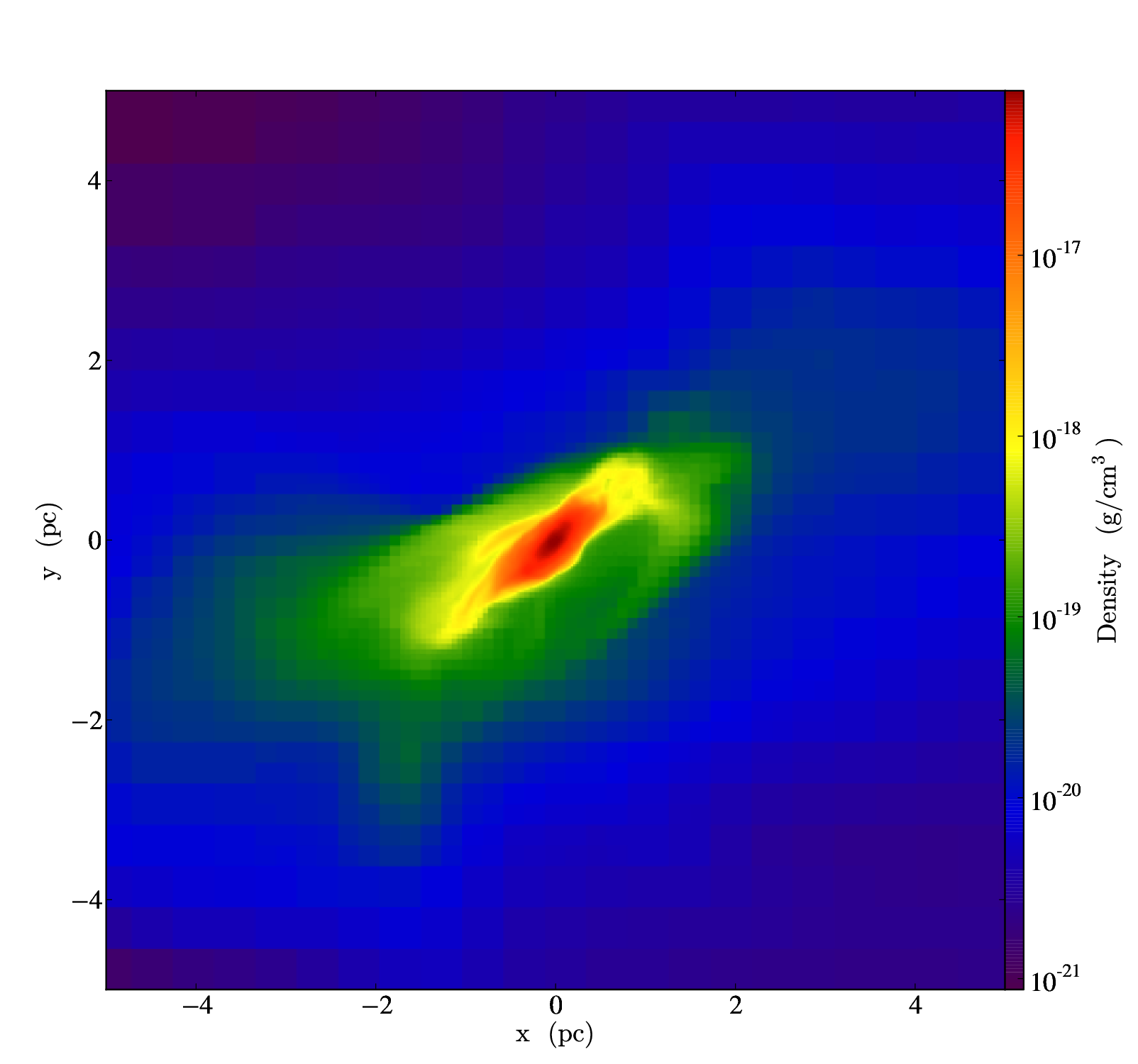}
\end{minipage} 
\end{tabular}
\caption{Density projections are shown for the central 10 pc for the run without sinks. The left and right panels show the projections along the y- and z-axis.}
\label{fig10}
\end{figure*}

The state of the simulations after twenty thousand years of the evolution is shown in figure \ref{fig10}. It is clearly visible from the density projections that a disk is formed in the center of the halo. The mass of the disk is  $\rm \sim 10^{5}~M_{\odot}$ equivalent to the sink mass in the corresponding run.  The formation of a parsec size disk is according to the expectation of theoretical models for the black hole formation \citep{2006MNRAS.371.1813L}. Both approaches provide high masses in relatively short time scales. This verification of results without sinks puts our estimates on even sounder footing.

\begin{figure*}
\centering
\begin{tabular}{c c}
\begin{minipage}{4cm}
\hspace{-4cm}
\includegraphics[scale=0.38]{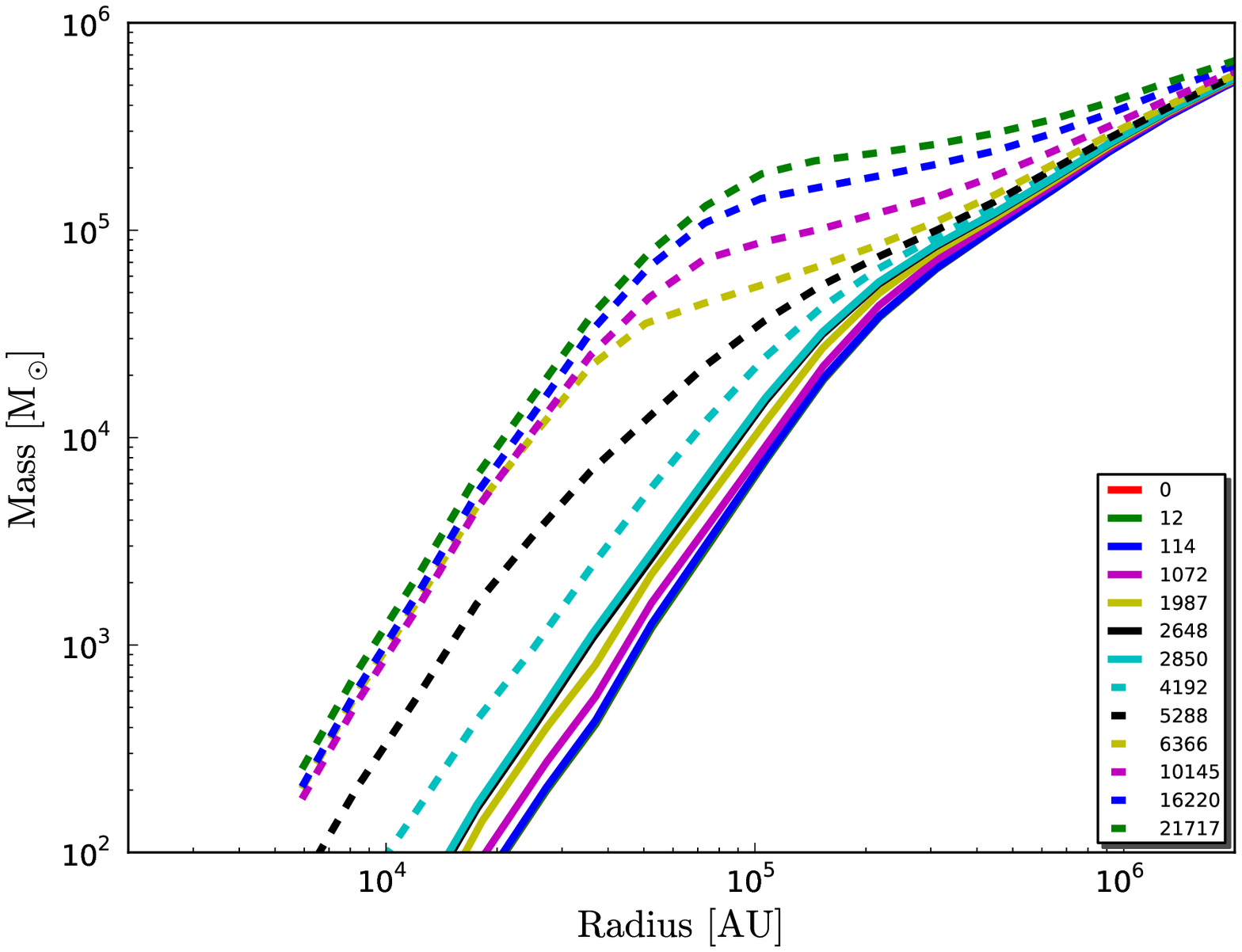}
\end{minipage} &
\begin{minipage}{4cm}
\includegraphics[scale=0.38]{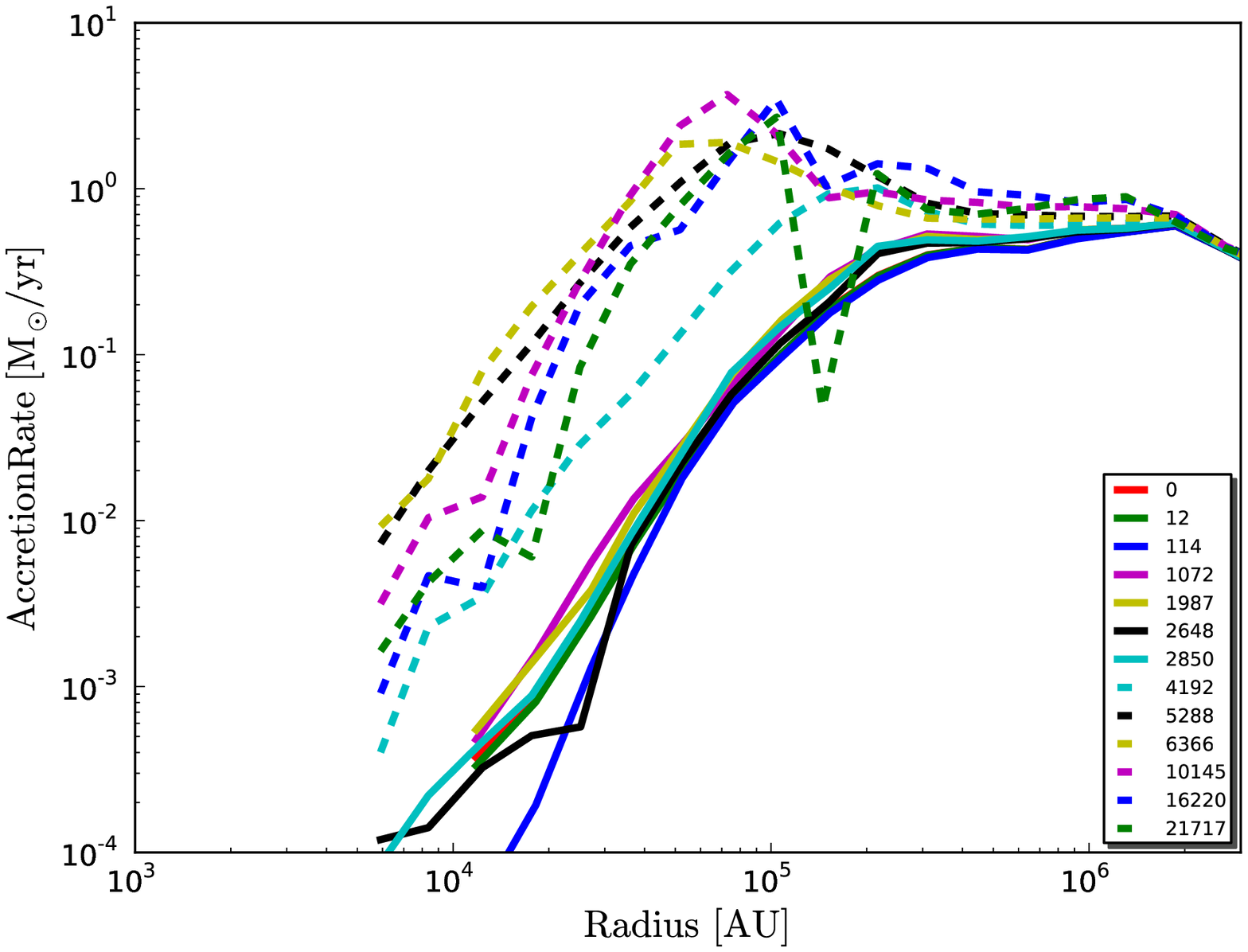}
\end{minipage} 
\end{tabular}
\caption{Time evolution of the mass and accretion rate radial profiles for the run without sinks. The left panel shows the mass radial profiles and the right panel shows the accretion rate radial profiles. The time corresponding to each radial profile is shown in the legend and its units are in years.} 
\label{fig11}
\end{figure*}

We further show the time evolution of the mass radial profile for this run as depicted in the left panel of figure \ref{fig11}. It can be noticed that initially the mass follows  an $\rm r^2$ behavior within the Jeans length and then increases linearly with radius. This trend is consistent with an isothermal density profile. Over the passage of time, the mass increases due to the infall of gas in the center of the halo and profile gets flattened. A disk of $\rm \sim 10^{5}~M_{\odot}$ is formed and most of the mass lies in the centeral clump as indicated by the flat mass profile. The time evolution of the accretion rates radial profile is shown in the right panel of figure \ref{fig11}. Similar to sinks simulation, accretion rate increases with time and reaches a few solar masses per year in about 10,000 years.  We note that the accretion rate measured here effectively probes a larger scale, and the profiles indicate an increase towards smaller scales. We therefore consider them to be consistent with sink particle runs.

\section{Discussion}

We present here the results from the first cosmological large eddy simulations employing sink particles and following the collapse for 20,000 years after their formation. These simulations are performed for three distinct halos and the results are compared with implicit large eddy simulations. The main objective of this study is to compute the characteristic mass scale of seed black holes resulting from the direct collapse model. We also computed the time evolution of mass accretion rates in massive primordial halos irradiated by a strong Lyman Werner UV background flux. 

Our findings show that black hole seeds with characteristic masses of $\rm 10^{5}~M_{\odot}$ are formed in a short time scale of twenty thousand years after their formation. It is further found that the characteristic masses are two times higher in LES. The time evolution of the accretion rates shows a characteristic behavior, it increases with time and reaches a peak value of $\rm 10~M_{\odot}/yr$. The accretion rate becomes almost constant as the rotational support is increased in the halo at later times. We further noticed that multiple sinks are formed in one halo with masses between $\rm 5 \times 10^{4}-10^{5}~M_{\odot}$. It is worth mentioning that we confirmed our estimates for the masses of sinks by evolving simulations adiabatically soon after they reached the maximum refinement level (an alternative approach to sinks) and found similar results. We further stress that our estimates for the characteristic mass are robust as they are confirmed from two independent approaches. 

The results from this study suggest that the formation of supermassive stars of $\rm 10^5~M_{\odot}$ seems the most plausible outcome as an intermediate stage to the formation of supermassive black holes. This is in accordance with the prediction of theoretical studies \citep{2010MNRAS.402..673B,2012ApJ...756...93H,2013arXiv1305.5923S,2013arXiv1308.4457H}. We have further shown that higher accretion rates of $\rm \geq 0.1~M_{\odot}/yr$ can be maintained for longer time scales. Our simulations show that SGS turbulence favors higher accretion rates compared to the ILES and the resulting seed black holes are about two times more massive.

It is expected that supermassive protostars produce accretion luminosity feedback during their early stage. Our calculations did not take into account this effect. We expect that the accretion luminosity feedback will have only minor impact on the masses of seed black holes as similar study exploring this impact in minihalos reported no significant impact \citep{2011MNRAS.414.3633S,2012MNRAS.424..457S}. The UV feedback is expected to occur when the star mass exceeds $\rm 10^{5}~M_{\odot}$ \citep{2013arXiv1308.4457H} which may further influence the growth of such stars \citep{2012ApJ...756...93H,2012ApJ...750...66J}. Numerical simulations exploring the impact of UV feedback should be performed in the future. 

\section*{Acknowledgments}
The simulations described in this work were performed using the Enzo code, developed by the Laboratory for Computational Astrophysics at the University of California in San Diego (http://lca.ucsd.edu). We acknowledge research funding by Deutsche Forschungsgemeinschaft (DFG) under grant SFB $\rm 963/1$, projects A12, A15 and computing time from HLRN under project nip00029. DRGS thanks the DFG for funding via the Schwerpunktprogram SPP 1573 ``Physics of the Interstellar Medium'' (grant SCHL $\rm 1964/1-1$). The simulation results are analyzed using the visualization toolkit for astrophysical data YT \citep{2011ApJS..192....9T}.

\bibliography{sink.bib}

\end{document}